\documentclass[12pt]{article}
\usepackage{latexsym}
\usepackage{amsmath,,calrsfs}
\usepackage{amsfonts}
\usepackage{amssymb}
\usepackage{amscd}
\usepackage{fancybox}
\usepackage{cite}
\usepackage{amsmath,amsfonts,amsbsy}
\usepackage{pstricks,pst-node}
\usepackage[small,bf,hang]{caption2}

\usepackage[vcentermath]{youngtab}

\usepackage{float}

\psset{unit=1.3cm,linewidth=.5pt,radius=.2}  

\newcommand{\fund}{{\tiny \yng(1)}}
\newcommand{\twoform}{{\tiny \yng(1,1)}}

\newcommand{\fourform}{{\tiny \yng(1,1,1,1)}}

\newcommand{\symm}{{\tiny \yng(2)}}

\newcommand{\block}{{\tiny \yng(2,2)}}
\newcommand{\foursymm}{{\tiny \yng(4)}}

\usepackage[vcentermath]{youngtab}          
\usepackage{enumitem}                       
\usepackage{multirow}                     
\usepackage{float}                          
\usepackage{lscape}                         
\usepackage{bm}

\def\hybrid{\topmargin -20pt    \oddsidemargin 0pt
        \headheight 0pt \headsep 0pt
        \textwidth 6.25in       
        \textheight 9.5in       
        \marginparwidth .875in
        \parskip 5pt plus 1pt   \jot = 1.5ex}

\hybrid

\newcommand{\mN}{{\mathcal N}}

\newcommand{\cL}{{\cal L}}
\newcommand{\cM}{{\cal M}}
\newcommand{\cN}{{\cal N}}

\newcommand{\beq}{\begin{equation}}
\newcommand{\eeq}{\end{equation}}
\newcommand{\bi}{\begin{itemize}}
\newcommand{\ei}{\end{itemize}}
\newcommand{\bea}{\begin{eqnarray}}
\newcommand{\eea}{\end{eqnarray}}
\newcommand{\ba}{\begin{array}}
\newcommand{\ea}{\end{array}}
\newcommand{\bt}{\begin{tabular}}
\newcommand{\et}{\end{tabular}}
\newcommand{\bc}{\begin{center}}
\newcommand{\ec}{\end{center}}

\def\theequation{\arabic{section}.\arabic{equation}}

\newcommand{\ft}[2]{{\textstyle {\frac{#1}{#2}} }}

\begin{document}

\begin{titlepage}
\begin{center}

\hfill UG-08-11 \\
\hfill UB-ECM-PF-08/16\\
\hfill ENSL-00299637\\
\hfill MIFP-08-18\\
\vskip 1cm

{\Large \bf The Superconformal Gaugings in Three Dimensions} \\[0.2cm]

\vskip 1cm

{\bf Eric A.~Bergshoeff\,$^\dagger$, Olaf Hohm\,$^\dagger$, Diederik
Roest\,$^*$,\\
  Henning Samtleben\,$^{**}$ and Ergin
Sezgin\,$^\ddagger$}\footnote{e-mail:~e.a.bergshoeff@rug.nl,
  o.hohm@rug.nl, droest@ecm.ub.es, henning.samtleben@ens-lyon.fr,
sezgin@physics.tamu.edu} \\

\vskip 25pt

$^\dagger$  {\em Centre for Theoretical Physics, University of
Groningen, \\
Nijenborgh 4, 9747 AG Groningen, The Netherlands \vskip 5pt }


\vskip 15pt

$^*$ \hskip -.1truecm {\em Departament Estructura i Constituents de la Materia \\
    \& Institut de Ci\`{e}ncies del Cosmos, \\
    Universitat de Barcelona \\
Diagonal 647, 08028 Barcelona, Spain \vskip 5pt }


\vskip 15pt

$^{**}$ {\em Universit\'e de Lyon, Laboratoire de Physique,\\
Ecole Normale Sup\'erieure de Lyon,\\
46 all\'ee dItalie, F-69364 Lyon Cedex 07, France \vskip 5pt}


\vskip 15pt

$^\ddagger$ {\em George P. and Cynthia W. Mitchell Institute for
Fundamental Physics,\\
Texas A\&M University, College Station, TX 77843-4242, U.S.A \vskip
5pt}


\vskip 0.8cm

\end{center}

\vskip 1cm

\begin{center} {\bf ABSTRACT}\\[3ex]

\begin{minipage}{13cm}
\small

We show how three-dimensional superconformal theories for any number
${\cal N}\le 8$ of supersymmetries can be obtained by taking a
conformal limit of the corresponding three-dimensional gauged
supergravity models. The superconformal theories are characterized
by an embedding tensor that satisfies a linear and quadratic
constraint. We analyze these constraints and give the general
solutions for all cases. We find  new  ${\cal N}=4,5$ superconformal
theories based on the exceptional Lie superalgebras $F(4), G(3)$ and
$D(2|1;\alpha)$. Using the supergravity connection we discuss which
massive deformations to expect. As an example we work out the
details for the case of ${\cal N}=6$ supersymmetry.

\end{minipage}

\end{center}


\vfill

July 2008

\end{titlepage}

\section{Introduction}\setcounter{equation}{0}

Three-dimensional superconformal theories have been studied
intensively recently in view of their relevance to describing the
dynamics of multiple M2-branes \cite{Schwarz:2004yj}. The starting
point of this development was the construction of the ${\cal N}=8$
supersymmetric models by Bagger and Lambert
\cite{Bagger:2006sk,Bagger:2007jr,Bagger:2007vi} and Gustavsson
\cite{Gustavsson:2007vu,Gustavsson:2008dy}. These models are
invariant under the symmetries of the ${\rm OSp}(8|4)$
superconformal algebra \cite{Bandres:2008vf}. However, under certain
assumptions, there is only a single ${\cal N}=8$ model with ${\rm
SO}(4)$ gauging \cite{Papadopoulos:2008sk,Gauntlett:2008uf}.

A way to obtain more general gauge groups is to consider models with
less superconformal symmetry, like  ${\cal N}=1,2$
\cite{Zupnik:1988en,Ivanov:1991fn,Avdeev:1991za,Gaiotto:2007qi},
${\cal N}=3$ \cite{Zupnik:1988wa,Kao:1993gs} or ${\cal N}=4$
\cite{Gaiotto:2008sd,Fuji:2008yj,Hosomichi:2008jd}. A particularly
interesting class of models are the  ${\cal N}=6$ models
 with ${\rm SU}(N)\times {\rm SU}(N)$ gauge
groups \cite{Aharony:2008ug}. Recently, three papers have appeared
that deal with the general construction of  ${\cal N}=6$
superconformal theories. First of all, in \cite{Hosomichi:2008jb} ${\cal
N}=6$ superconformal  models were constructed starting from ${\cal
N}=4$ supersymmetry with special matter multiplets and making use of
a relation with Lie superalgebras \cite{Gaiotto:2008sd}.  Secondly, in
\cite{Bagger:2008se} a general framework for constructing ${\cal
N}=6$ superconformal gauge theories using the three-algebra approach
was presented. Finally, a group-theoretical classification of the
gauge groups and matter content of ${\cal N}=6$ superconformal gauge
theories was given in \cite{Schnabl:2008wj}.

In this note we wish to approach the construction of
three-dimensional superconformal gauge theories for all values
of ${\cal N}$ by making use of a relation with  gauged supergravity
\cite{Bergshoeff:2008cz,Bergshoeff:2008ix}. Three-dimensional gauged
supergravities have been constructed  using the so-called embedding
tensor technique. This method
 was originally developed in the construction of maximal ${\cal
N}=16$ supergravities \cite{Nicolai:2000sc,Nicolai:2001sv}, where the most general  ${\cal N}=16$  gaugings encoded in the  ``embedding tensor'' were classified.
The role of this tensor is to specify which
subgroup of the global symmetry group is gauged and which vectors
are needed to perform this gauging. Later, the same technique was
applied to construct the matter-coupled half-maximal ${\cal N}=8$
theory \cite{Nicolai:2001ac,deWit:2003ja} as well as the ${\cal N}<
8$ theories \cite{deWit:2003ja}.

In  \cite{Bergshoeff:2008ix} it was shown how the ${\cal N}=8$ model
of \cite{Bagger:2006sk,Bagger:2007jr,Bagger:2007vi,Gustavsson:2007vu,Gustavsson:2008dy}
can be obtained by taking an appropriate limit to global
supersymmetry of the ${\cal N}=8$ supergravity model of
\cite{Nicolai:2001ac,deWit:2003ja}. We will refer to this limit as
the conformal limit. We  consider here the conformal limit of all
the other gauged supergravity  models. In general, the embedding
tensor characterizing the superconformal theory satisfies a set of
linear and quadratic constraints. We  show how these constraints can
be determined from gauged supergravity by taking the conformal limit
and find agreement with all known structures. Furthermore, we
present a systematic way to solve these constraints, which
reproduces the classification of superconformal theories for
different values of ${\cal N}$ given in the recent literature
\cite{Gaiotto:2008sd,Hosomichi:2008jd,Hosomichi:2008jb,Schnabl:2008wj}.
Furthermore, inspired by a connection between superconformal gauge
theories and Lie superalgebras observed in \cite{Gaiotto:2008sd}, we
 construct new ${\cal N}=4,5$ superconformal theories that are
based on the exceptional Lie superalgebras $F(4), G(3)$ and
$D(2|1;\alpha)$ with $\alpha$ a free parameter. They lead to
superconformal theories with ${\rm SO}(7)\times {\rm Sp}(1),
G_2\times {\rm Sp}(1)$ and ${\rm SO}(4)\times {\rm Sp}(1)$ gaugings,
respectively\footnote{We thank J. Park for a stimulating discussion
regarding the case of $F(4)$.}.

One advantage of the supergravity approach is that the same idea can
be used to obtain  non-conformal theories as well  by taking other
limits. A particularly interesting class of models is obtained by
taking the limit of a gauged supergravity where the gauge group lies
entirely within the $R$-symmetry group. As was shown in
\cite{Bergshoeff:2008ix} for the $\cN = 8$ case, such gaugings do
not survive the limit to global conformal supersymmetry but
nevertheless give rise to massive deformations instead. These are
important to test the idea of multiple M-branes and have been
considered for ${\cal N}=8$ \cite{Gomis:2008cv,Hosomichi:2008qk} and
${\cal N}=6$ \cite{Hosomichi:2008jb,Gomis:2008vc,Hanaki:2008cu}.

This work is organized as follows. In section 2 we discuss the
conformal limit and show how the properties of the embedding tensor
characterizing superconformal theories in three dimensions can be
derived from gauged supergravity. In particular, we derive the
linear and quadratic constraints these tensors must satisfy for different values of $\cN$. In
section 3 we perform a systematic analysis of these constraints and
derive the possible gauge groups and matter content. Next, we work out the details of our
method for the specific example of ${\cal N}=6$ supersymmetry in
section 4. Finally, in section 5 we present our conclusions. In
particular, we  comment on the possible massive deformations.
Appendix A explains our notation and conventions.

\section{Superconformal gaugings in three dimensions} \setcounter{equation}{0}

\subsection{Gauged supergravity}

We begin with a review of the possible gauged supergravity theories
in three dimensions with different numbers ${\cal N}$ of
supersymmetries. We are interested in theories with $\mN \le 8$ as
these have matter multiplets (in addition to the supergravity
multiplet) and allow for a limit to a globally supersymmetric field
theory.

Three-dimensional supergravity theories differ from their
higher-dimensional relatives in that all bosonic degrees of freedom
can be described by scalar fields. These can be seen as coordinates
of a manifold, on which supersymmetry imposes a number of geometric
conditions \cite{Tollsten}. For $\mN > 4$ these are strong enough to
completely fix the (ungauged) theory: the scalar manifolds are given
by certain symmetric spaces of the form
 \begin{align}
  \cM = \frac{\hat G}{\hat H} \,, \label{coset}
 \end{align}
where $\hat G$ is a simple Lie group of isometries, and $\hat H$ is
its maximal compact subgroup. For lower values of $\mN$, the scalar
manifolds can be more general manifolds such as quaternionic,
K\"ahler and Riemannian manifolds. However, for our purposes it will
be sufficient to consider certain symmetric spaces for $\mN \leq 4$
as well. The different cases are summarised in table
\ref{tab:manifolds}. Note that the $\mN=4$ scalar manifold consists
of a product of two quaternionic spaces. This possibility occurs due
to the existence of two inequivalent $\mN=4$ matter multiplets,
hyper and twisted hyper multiplets. For other values of $\mN$ there
is a unique matter multiplet.

\renewcommand{\arraystretch}{1.4}
\begin{table}[ht]
\begin{center}
$
\begin{array}{||c||c|c||c||}
\hline \mN &  \hat G & \hat H & {\rm dim}[\hat G / \hat H] \\ \hline \hline
8 & {\rm SO}(8,N) & {{\rm SO}(8)\times {\rm SO}(N)} & 8N \\[1mm] \hline
6 & {\rm SU}(4,N) & {{\rm S(U}(4)\times {\rm U}(N))} & 8N \\[1mm]\hline
5 & {\rm Sp}(2,N) & {{\rm Sp}(2)\times { \rm Sp}(N)} & 8N \\[1mm]\hline
4 & {\rm Sp}(1,N) \times {\rm Sp}{(1,N')} & {\rm Sp}(1 ) \times {\rm Sp}(N) \times {\rm Sp}(1) \times {\rm Sp}(N') & 4N + 4N' \\[1mm] \hline
3 & {\rm Sp(1,N)} & {\rm Sp}{(1)\times {\rm Sp}(N)} & 4N \\[1mm] \hline
2 & {\rm SU}(1,N) & {{\rm S(U}(1) \times {\rm U}(N))} & 2N \\[1mm] \hline
1 & {\rm SO}(1,N) & {{\rm SO}(N)} & N \\[1mm]   \hline
\end{array}$
\caption{The isometry and isotropy groups $\hat G$ and $\hat H$ of
the symmetric scalar manifolds of three-dimensional $\mN$-extended
supergravity and their dimensions. For $\cN \le 4$ we have just
included a particular series of symmetric spaces, as it turns out
that these contain the most general global limit to flat target
spaces.} \label{tab:manifolds}
\end{center}
\end{table}
\renewcommand{\arraystretch}{1}

Turning to gauged supergravity, it is important to note that a
special $D=3$ feature is that the gauge vectors have no independent
kinetic term but only occur via a Chern-Simons term. In this way
they do not introduce new degrees of freedom but are dual to the
scalar fields. More precisely, one can introduce as many vector
fields as there are isometries on the scalar target space.

The possible gaugings of $D=3$ supergravity theories have been
classified using the embedding tensor approach \cite{Nicolai:2000sc,
Nicolai:2001sv,deWit:2003ja}. The embedding tensor $\Theta_{\alpha
\beta} = \Theta_{ \beta\alpha}$ takes values in the symmetric
product of two adjoint representations of the global symmetry group
$\hat{G}$:
 \begin{align}
   \Theta \in ({\rm Adj}(\hat G) \otimes {\rm Adj}(\hat G))_{\rm symm} \,,
 \end{align}
and relates gauge vectors to generators of $\hat{G}$. The associated
transformations are then gauged due to the introduction of the
embedding tensor in covariant derivatives which take the general
form
\begin{equation}
D_\mu = \partial_\mu - A_\mu{}^\alpha\, \Theta_{\alpha\beta}\,
t^\beta\,,
\end{equation}
for some representation matrices $t^\beta$ of $\hat{G}$. Note that
$X_\alpha = \Theta_{\alpha\beta}t^\beta$ denote the  generators
whose symmetries are being gauged. Also, the embedding tensor
appears as a metric in the Chern-Simons term
\begin{equation}
{\cal L}_{\rm CS} =
-\ft12\varepsilon^{\mu\nu\rho}A_{\mu}{}^{\alpha}\Theta_{\alpha\beta}\left(\partial_{\nu}A_{\rho}{}^{\beta}
   -\ft13\Theta_{\gamma\delta}f^{\beta\delta}{}_{\epsilon}A_{\nu}{}^{\gamma}A_{\rho}{}^{\epsilon}\right) \;, \label{LCS}
\end{equation}
with vector fields transforming in the adjoint of $\hat{G}$ and the
structure constants $f^{\alpha\beta}{}_\gamma$ of the global
symmetry group $\hat{G}$. In supergravity there is a number of
restrictions on which transformations can be gauged. These can be
succinctly summarised in terms of a linear and a quadratic
constraint on the embedding tensor.

The quadratic constraint follows from the requirement that the embedding tensor itself is
invariant under the transformations that are gauged due to the introduction of $\Theta$.
This condition takes the same form for all values of $\mN$:
 \bea \label{quadraticall}
  \Theta_{\alpha \beta} \Theta_{\gamma ( \delta}\,f^{\alpha \gamma}{}_{\epsilon)} & = &0
  \,.
\label{quadratic} \eea
In case the embedding tensor projects onto a
semisimple subgroup of $\hat{G}$ and is expressed in terms of
invariant tensors of that subgroup, the quadratic
constraint~(\ref{quadratic}) is automatically satisfied.

The linear constraint on the embedding tensor follows from supersymmetry. In other words, it is perfectly consistent to introduce
gaugings that do not satisfy the linear constraint, but these will not preserve supersymmetry. As it follows from the requirement
of supersymmetry, this condition takes a different form for different values of $\mN$:

\bigskip

\noindent $\bullet\ $\underline{$\mN =8$ supergravity:} \\[.1cm] \noindent
The embedding tensor takes values in the symmetric product of the
adjoint of
   $\hat G = {\rm SO}(8,N)$. Representing the adjoint index $\alpha$ by a pair of antisymmetric fundamental
   indices, i.e.~$\alpha = [AB]$, the embedding tensor is of the
form $\Theta_{[AB],[CD]}$ and in terms of ${\rm SO}(8,N)$ Young
tableaux\footnote{Here, we use Young tableaux of SO(8,N) in which
symmetrization refers to traceless symmetrization, such that the
representations are irreducible.}
 decomposes according to
    \begin{align}
     (\twoform \otimes \twoform)_{\rm symm} = 1 \oplus \symm \oplus \fourform \oplus \block \,. \label{N=8}
    \end{align}
    Supersymmetry requires absence of the last representation corresponding to the window tableau.

   \bigskip

\noindent $\bullet\ $\underline{$\mN = 6$ supergravity:} \\[.1cm] \noindent The embedding tensor takes values in the symmetric product of the
   adjoint of $\hat G = {\rm SU}(4,N)$. Representing the adjoint index $\alpha$
   by a lower fundamental index $A$ and an upper anti-fundamental
   index $B$, i.e.~$\alpha=({}_A,{}^B)$, the embedding tensor is of the
form $\Theta_{A}{}^B{},_C{}^D$ and in terms of ${\rm SU}(4,N)$ Young
tableaux decomposes according to\,\footnote{We use here a notation
where a
   (barred) Young tableau
denotes
   (upper) lower indices of a tensor and traces are
subtracted.}:
    \begin{align}
     (\fund \, \overline \fund \otimes \fund \, \overline \fund)_{\rm symm} = 1 \oplus \fund \, \overline \fund \oplus \twoform \, \overline \twoform
     \oplus \symm \, \overline \symm \,.
    \end{align}
   As supersymmetry requires the last representation to vanish, the embedding tensor is anti-symmetric
 in its lower two indices: $\Theta_A{}^B{}_{,\,C}{}^D = \Theta_{[A}{}^B{}_{,\,C]}{}^{D}$.

    \bigskip

\noindent $\bullet\ $\underline{$\mN = 5$ supergravity:} \\[.1cm] \noindent The embedding tensor takes values in the symmetric product of the adjoint of
   $\hat G = {\rm Sp}(2,N)$. Representing the adjoint index $\alpha$ by a pair of symmetric fundamental
   indices,
   i.e.~$\alpha = (AB)$,  the embedding tensor is of the
form $\Theta_{(AB),(CD)}$ and in terms of ${\rm Sp}(2,N)$ Young
tableaux decomposes according to
    \begin{align}
     (\symm \otimes \symm)_{\rm symm} = 1 \oplus \twoform \oplus \block \oplus \foursymm \,.
    \end{align}
  The last representation again has to be absent for supersymmetric gaugings, leading to the following  constraint on the embedding tensor:
   $\Theta_{(AB,CD)} = 0$.

    \bigskip

\noindent $\bullet\ $\underline{$\mN = 4$ supergravity:} \\[.1cm] \noindent In this case the embedding tensor consists of three parts.
  One part, $\Theta_{AB,CD}$, takes values in the symmetric product of the adjoint of
  ${\rm Sp}(1,N)$ and
  satisfies the analogous conditions as for $\mN = 5$. Similarly, the other part
  $\Theta_{A'B',C'D'}$ takes values in the symmetric product of the adjoint of ${\rm Sp}(1,N')$ and is subject to the same conditions.
  The last part, $\Theta_{AB,C'D'}$
  takes values in the product of the adjoints of both factors of the global symmetry group
  $\hat G$ and is not subjected to any linear constraint. However, it does have to
  satisfy a quadratic condition:
 \begin{align} \label{quadraticmix}
  \Theta_{\alpha \beta} \Theta_{\gamma \epsilon'} f^{\alpha
\gamma}{}_\delta + \Theta_{\alpha' \beta} \Theta_{\gamma' \delta}
f^{\alpha' \gamma'}{}_{\epsilon'} = 0 \,,
 \end{align}
and similar for $(\alpha \leftrightarrow \alpha')$.

    \bigskip

\noindent $\bullet\ $\underline{$\mN \leq 3$ supergravity:} \\[.1cm] \noindent In these cases supersymmetry does not impose any linear condition: all consistent gaugings (satisfying the quadratic constraint) are compatible with supersymmetry. In addition, for $\cN = 1,2$, there are deformations that do not correspond to any gauging but to the introduction of a superpotential instead \cite{deWit:2003ja}.

There are in general two strategies to solve the set of linear and
quadratic constraints on the embedding tensor. Either one starts
from an embedding tensor which projects onto a given subgroup by
means of an invariant tensor such that the quadratic constraint is
automatically satisfied. In this case, the linear constraint becomes
a non-trivial identity which decides if the gauging is a viable one.
Alternatively, one may start from the general solution of the linear
constraint which can directly be expressed in terms of the proper
subrepresentations. Then, the quadratic constraint becomes a
non-trivial identity which selects the proper gaugings. In both
cases the embedding tensor represents  the Cartan-Killing metric of
the gauge group (at least for semi-simple gaugings
\cite{deWit:2003ja}). However, while in solving the quadratic
constraint first, we can specify the Cartan-Killing metric in a
preferred basis, e.g.~the diagonal one, if we solve the linear
constraint first instead, the subsequent solution of the quadratic
constraint, if it exists at all, yields the Cartan-Killing metric in
a particular basis over which we no longer have control, e.g.~it may
emerge in a non-diagonal basis.

  \bigskip

\subsection{The conformal limit}

We now turn to the conformal limit, whose aim is to extract superconformal
theories from the gauged supergravities discussed above.
This limit was performed explicitly for $\cN = 8$ in \cite{Bergshoeff:2008ix}
and will be generalised here to lower values of $\cN$.

It will be instructive to first discuss the limit to global
supersymmetry in the ungauged case. Upon sending  Newton's constant
to zero, the supergravity and matter multiplets decouple, and the
former will be set to zero. The resulting theory for the matter
multiplets has $\cN$ global supersymmetries. The isometry groups of
supergravity, see table \ref{tab:manifolds}, can be seen to split up
into three parts. Its compact part, which is the product of the
R-symmetry group $H_{\rm R} = {\rm SO}(\cN)$ and its orthogonal
complement $G$, are unaffected by the global limit. In contrast, the
non-compact generators reduce to nilpotent generators that transform
under the compact parts:
 \begin{align}
  \hat G \rightarrow  (G \times H_{\rm R}) \ltimes \mathbb{R}^{cN} \,,
 \end{align}
for integer $N$ and where $c=1,2,4$ or $8$ depending on the size of
the matter multiplet of $\cN$-extended supersymmetry. The resulting
groups are summarised in table \ref{tab:groups}.

\renewcommand{\arraystretch}{1.4}
\begin{table}[ht]
\begin{center}
$
\begin{array}{||c||c|c||c||}
\hline \mN &  G & H_{\rm R} & {\rm dim} [\mathbb{R}^{cN}] \\ \hline
\hline
8 & {\rm SO}(N) & {{\rm SO}(8)} & 8N \\[1mm] \hline
6 & {\rm U}(N) & {{\rm SU}(4)} & 8N \\[1mm]\hline
5 & {\rm Sp}(N) & {{\rm Sp}(2)} & 8N \\[1mm]\hline
4 & {\rm Sp}(N) \times {\rm Sp(N')} & {\rm Sp}(1 )\times {\rm Sp}(1) & 4N + 4N' \\[1mm] \hline
3 & {\rm Sp(N)} & {\rm Sp(1)} & 4N \\[1mm] \hline
2 & {\rm SU}(N) & {{\rm U}(1)} & 2N \\[1mm] \hline
1 & {\rm SO}(N) & {\rm 1} & N \\[1mm]   \hline
\end{array}$
\caption{The global symmety and R-symmetry groups $G$ and $H_{\rm R}$ of three-dimensional $\mN$-extended field theory and the dimension of the flat scalar manifolds.} \label{tab:groups}
\end{center}
\end{table}
\renewcommand{\arraystretch}{1}

Our notation is as follows: the fundamental representation of $\hat
G$ splits up according to $A = (I,a)$, where $I$ is the fundamental
representation of the R-symmetry group $H_{\rm R}$ and $a$ of the
global symmetry group $G$. The scalar fields correspond to the
non-compact generators and are denoted by $X^{Ia}$. As these
correspond to nilpotent generators, the associated scalar manifolds
are flat in all cases we consider and the group $G$ acts as their
global symmetry. For $\cN = 4$ one gets two copies of flat
manifolds, composed of the two different types of matter multiplets.

In addition to ungauged theories with global supersymmetry, one can
also obtain their gauged counterparts from supergravity. As shown in
\cite{Bergshoeff:2008ix}, one can derive $\cN = 8$ conformal as well
as non-conformal gaugings and even massive deformations from the
corresponding supergravity by taking the proper global limit. In the
case of conformal gaugings, it can be seen that this requires the
embedding tensor $\Theta$ to be a singlet of the R-symmetry group,
i.e.~to only take values in the symmetric product of the adjoint of
the global symmetry groups $G$. Only the components $\Theta_{ab,cd}$
lead to a conformal gauging, while other components can lead to
non-conformal gaugings  and/or massive deformations. Indeed, in the
analysis of \cite{Bergshoeff:2008ix} it was found that the other
components of the embedding tensor have to be rescaled with Newton's
constant in order to avoid singular terms. This rescaling changes
the mass dimension of these components, such that they have exactly
the mass dimension of non-conformal gaugings and/or massive
deformations. The components $\Theta_{ab,cd}$ that we will retain do
not require such a rescaling and indeed correspond to conformal
gaugings.

It remains to be seen which components $\Theta_{ab,cd}$ can be
obtained from gauged supergravity. In \cite{Bergshoeff:2008ix} it
was found for ${\cal N}=8$ that only the four-form representation in
\eqref{N=8} gives rise to conformal gaugings. The other
representations in \eqref{N=8} give rise to non-zero values for
other components of the embedding tensor, involving the $R$-symmetry
directions, and hence they spoil the conformal invariance. Therefore
for $\cN = 8$ globally supersymmetric field theories we obtain
conformal gaugings parametrised by an embedding tensor in the
four-form representation of the global symmetry group $G$, which is
${\rm SO}(N)$ for the case of global supersymmetry. This is in
precise agreement with the findings in the direct
construction~\cite{Bagger:2006sk,Bagger:2007jr,Bagger:2007vi,Bergshoeff:2008cz}.

A short analysis reveals that the situation is slightly different
for the theories with less than ${\cal N}=8$ supersymmetry, in that
one can use {\sl all} representations of the supergravity embedding
tensor to obtain conformal gaugings\,\footnote{The reason is that
for lower ${\cal N}$, the corresponding components in the embedding
tensor can be excited independently without inducing components in
other blocks of the embedding tensor, see the appendix
of~\cite{deWit:2003ja} for the detailed decompositions. For ${\cal
N}=8$ in contrast, a non-vanishing component $\symm$ within $SO(N)$
induces components in the non-compact part of the embedding tensor
which spoil the global limit.}. These are therefore classified by
exactly the same representations that solve the linear constraint in
the supergravity case, except that the linear constraint now refers
to $G$ instead of $\hat G$. Similarly, the quadratic constraint
takes the same form \eqref{quadraticall}, but where the structure
constants refer to $G$ instead of $\hat G$ as well. In more detail,
we find the following conditions for the different cases:

    \bigskip

\noindent $\bullet\ $\underline{$\mN =8$ field theory:} \\[.1cm] \noindent The embedding tensor takes values in the following representation of $G= {\rm SO}(N)$:
    \begin{align}
     \fourform \,,
    \end{align}
    and as a consequence is totally anti-symmetric
    \begin{equation}\label{linear8}\Theta_{ab,cd} =
    \Theta_{[ab,cd]}\,.
    \end{equation}
With the ${\rm SO}(N)$ structure constants
 \begin{eqnarray}
 f^{ab,cd,}{}_{ef}  &=& -2\delta^{[a}{}_{[e}\delta^{b][c}\delta^{d]}{}_{f]}\;,
\end{eqnarray}
the quadratic constraint~(\ref{quadratic}) takes the explicit form
\begin{eqnarray}
\label{Q8} \Theta_{ab,e}{}^g\Theta_{cd,gf} +
\Theta_{ab,c}{}^g\Theta_{ef,gd}
-\Theta_{ab,f}{}^g\Theta_{cd,ge}-\Theta_{ab,d}{}^g\Theta_{ef,gc}  &
= & 0\;.
\end{eqnarray}

    \bigskip

\noindent $\bullet\ $\underline{$\mN = 6$ field theory:} \\[.1cm] \noindent The embedding tensor takes values in the following representations of $G = {\rm U}(N)$:
    \begin{align}
     1 \oplus \fund \, \overline \fund \oplus \twoform \, \overline \twoform
     \,,
    \end{align}
   and therefore it satisfies the linear constraint
   \begin{equation}\label{linearc}
   \Theta_{(a}{}^b{}_{,\,c)}{}^d =0\,.
   \end{equation}
With the ${\rm U}(N)$ structure constants
\begin{eqnarray}
  f_{a}{}^b{},\,_c{}^d,\,^{f}{}_e &=&
  i\left(\delta_{c}{}^{b}\delta_{a}{}^{f}\delta_{e}{}^{d}
  -\delta_{a}{}^{d}\delta_{c}{}^{f}\delta_{e}{}^{b}\right)
  \;,
\end{eqnarray}
the quadratic constraint~(\ref{quadratic}) takes the explicit form
\begin{eqnarray}
\label{Q6} \Theta_c{}^{g},_e{}^{f}\,\Theta_g{}^{d},_a{}^{b}
-\Theta_g{}^{d},_e{}^{f}\,\Theta_c{}^{g},_a{}^{b}
+\Theta_a{}^{g},_e{}^{f}\,\Theta_g{}^{b},_c{}^{d}
-\Theta_g{}^{b},_e{}^{f}\,\Theta_a{}^{g},_c{}^{d} &=& 0\;.
\end{eqnarray}

    \bigskip

\noindent $\bullet\ $\underline{$\mN = 5$ field theory:} \\[.1cm] \noindent The embedding tensor takes values in the following representations of $G = {\rm Sp}(N)$:
    \begin{align}
     1 \oplus \twoform \oplus \block \,,
     \label{N=5gl}
    \end{align}
  and hence satisfies the linear constraint
  \begin{equation}\label{linear5}\Theta_{(ab,cd)} = 0\,.
  \end{equation}
With the ${\rm Sp}(N)$ structure constants
 \begin{eqnarray}
 f^{ab,cd,}{}_{ef} &=& -2\delta^{(a}{}_{(e}\Omega^{b)(c}\delta^{d)}{}_{f)}\;,
\end{eqnarray}
the quadratic constraint~(\ref{quadratic}) takes the explicit form
\begin{eqnarray}
\label{Q4}
\Omega^{gh}\left(\Theta_{ab,eg}\Theta_{hf,cd}+\Theta_{ab,fg}\Theta_{he,cd}
   +\Theta_{ab,cg}\Theta_{hd,ef}+\Theta_{ab,dg}\Theta_{hc,ef}\right)
   \ = \ 0\;.
\end{eqnarray}

    \bigskip

\noindent $\bullet\ $\underline{$\mN = 4$ field theory:} \\[.1cm] \noindent As in the $\cN = 4$ supergravity case,
  the embedding tensor consists of three parts, $\Theta_{ab,cd}$, $\Theta_{a'b',c'd'}$ and
   $\Theta_{ab,c'd'}$ that take values in the products of the adjoints of ${\rm Sp}(N)$ and ${\rm Sp}(N')$.
  The former two consist of the same representations \eqref{N=5gl} as in the $\cN = 5$ case,
while the latter is unconstrained. The quadratic constraints
\eqref{quadraticall} and \eqref{quadraticmix} can
  also be written in a form analogous to \eqref{Q4} using the $Sp(N)$ and $Sp(N')$ structure constants.

    \bigskip

\noindent $\bullet\ $\underline{$\mN \leq 3$ field theory:} \\[.1cm] \noindent As in the $\cN \leq 3$ supergravities,
  the embedding tensor can take arbitrary values in the symmetric product of the adjoints of $G$.
  All consistent gaugings (satisfying the quadratic constraint) are compatible with supersymmetry.

  \bigskip

\noindent In this way we have obtained a classification of the
possible superconformal gaugings for different values of $\cN$ in a
uniform way, starting from the classification of the possible
gaugings of supergravity. Of course, one still needs to solve the
constraints for the embedding tensor. As we have discussed above,
there are in general two approaches to solve these constaints,
starting by either choosing for $\Theta$ the projector onto a
subgroup, or by solving the linear constraint on the embedding
tensor first. In both cases, one set of constraints is trivially
satisfied while the other one becomes a non-trivial identity.
Both approaches have been pursued in the literature and depending on
the point of view, the remaining constraint (which is the linear one
in the superpotential formalism
of~\cite{Gaiotto:2008sd,Hosomichi:2008jb} and the quadratic one in
the 3-algebra formalism
of~\cite{Bagger:2007jr,Gustavsson:2007vu,Bagger:2008se}) has been
referred to as the {\em fundamental identity}, respectively.

It is interesting to compare our results to those obtained recently
by the mechanism of supersymmetry enhancement. In this approach
one starts from the superconformal theories
with $\cN \leq 3$ for which there is no restriction on the gauge
group and the representation of the matter multiplets. In our
context, this corresponds to the absence of a linear constraint on
the embedding tensor. It was found that supersymmetry could be
enhanced to $\cN = 4$ by certain restrictions on the gauge group and
its representations \cite{Gaiotto:2008sd}. These correspond to the
linear constraint $\Theta_{(ab,cd)} = 0$. Subsequently, it was noted
that twisted hypermultiplets could be added and in fact are
necessary for further supersymmetry enhancement
\cite{Hosomichi:2008jd, Hosomichi:2008qk}. The untwisted and twisted
sector have to be taken identical to gain one further supersymmetry.
In our notation, this corresponds to the identification of the three
parts of the $\cN = 4$ embedding tensor leading to one $\cN = 5$
embedding tensor $\Theta_{ab,cd}$ subject to the same linear
constraint \eqref{linear5}. Yet further enhancement to $\cN = 6$ and $\cN = 8$ is
possible by restricting to embedding tensors that satisfy the
corresponding linear constraints \eqref{linearc} and \eqref{linear8}, respectively.

The same superconformal gaugings can  therefore be obtained in a
methodical way from two independent and rather different approaches.
In supergravity, the phenomenon of supersymmetry enhancement does
not exist: one can not adjust the couplings of e.g.~$\cN =3$
supergravity to obtain an $\cN = 4$ theory. This can for example be
seen from the different supergravity multiplets: the number of
gravitini is different for these theories. Nevertheless, the
classification of supergravity gaugings reduces in the conformal
limit to the same classification of superconformal gaugings
that has  been obtained from a global supersymmetry viewpoint. It is
interesting to see that the analogous results have been obtained
independently on the local and the global supersymmetry side. Using
the conformal limit these two approaches can be related.

\section{Solving the constraints} \setcounter{equation}{0}

In this section we will show how to solve systematically the linear
and quadratic constraints. We will first explain the general
strategy and next discuss the cases for different values of ${\cal
N}$ separately. Recently \cite{Hosomichi:2008jb}, a  classification
of the different superconformal theories has been given starting
from ${\cal N}=4$ supersymmetry with special matter multiplets and
making use of a relation with Lie superalgebras
\cite{Gaiotto:2008sd}. An alternative derivation for $\cN = 6$, of a
more group-theoretical nature, can be found in
\cite{Schnabl:2008wj}. Here we will use an approach directly based
on the embedding tensor. Based on the relation with Lie
superalgebras we will also uncover new $\cN = 4,5$ superconformal
gaugings that correspond to exceptional cases.

Our starting point is an embedding tensor that has only directions
in the global symmetry group $G$.  The purpose of the embedding
tensor is to project the Lie algebra generators  of the global
symmetry group onto the generators  of the subgroup which is gauged.
As explained in the previous section, this tensor must satisfy
certain linear and quadratic constraints. Our strategy is to start
from an embedding tensor which projects onto a given subgroup such
that the quadratic constraint is automatically satisfied. In this
case, the linear constraint becomes a non-trivial identity which
decides if the gauging is a viable one.

For the classical Lie groups we will use the standard invariant
tensors $\delta_{ab}= \delta_{ba}$ (orthogonal groups)\,,
$\delta_a{}^b$ (unitary groups) and $\Omega_{ab}=-\Omega_{ba}$
(symplectic groups). Here $\delta$ denotes the Kronecker delta and
$\Omega$ the anti-symmetric symplectic tensor with inverse tensor
$\Omega^{ab}$, i.e. $\Omega_{ac}\,\Omega^{bc} = \delta_a{}^b$.
Besides these tensors we will also use special invariant tensors in
the case of ${\rm SO}(4)$, ${\rm SO}(7)$ and $G_2$ which will lead
to the new ${\cal N}=4,5$ superconformal theories. Our first task is
to construct, using the invariant tensors, the operators that
project the Lie algebra generators  of the global symmetry group
onto the generators  of the subgroup which is gauged. Furthermore we
will also need the operators that project onto the singlet
representation. These operators will be the building blocks from
which we will construct the embedding tensor. In the case of the
classical orthogonal, unitary and symplectic groups these building
blocks are given by:
\begin{eqnarray}\label{po}
 \text{SO($N$) singlet:~~} && \delta_{ab}\,\delta_{cd}\,,\hskip 2truecm \text{SO($N$) adjoint:~~}
   \delta_{c[a}\,\delta_{b]d}\,,\nonumber\\
  \text{SU($N$) singlet:~~} && \delta_a{}^b\,\delta_c{}^d\,,\hskip 1.8truecm
\text{SU($N$) adjoint:~~} (\delta_c{}^b\,\delta_a{}^d - {1\over
N}\delta_a{}^b\,\delta_c{}^d)\,,\nonumber\\
  \text{Sp($N$) singlet:~~} &&
\Omega_{ab}\,\Omega_{cd}\,,\hskip 1.7truecm \text{Sp($N$)
adjoint:~~} \Omega_{c(a}\,\Omega_{b)d}\,.
\end{eqnarray}
For ${\rm SO}(4)$ there is
an additional operator that projects onto the adjoint representation
given by
\begin{equation}\label{specialN=4}
 \text{SO($4$) adjoint:~~}  \epsilon_{abcd}\,.
\end{equation}
This operator will be needed in the construction of the ${\cal N}=8$
and one of the exceptional ${\cal N}=4,5$  superconformal theories.

Typically we will need to split the index $a$ according to a pair of
indices $(i,\bar{i})$:
\begin{eqnarray}\label{split}
a&\rightarrow&(i,\bar{i})\hskip 1truecm  {\rm with}\hskip 1truecm
i=1, \dots, m\,;\ \bar i=1, \dots, n\,,
\end{eqnarray}
corresponding to a bi-fundamental representation. These cases will
be referred to as matrix models. Clearly, $n=1$ is a special case
for which the matrix reduces to a vector, and the indices $a$ and
$i$ coincide. In principle one could consider the index $a$ to
represent a sum of an arbitrary set of representations of the gauge
group other than those described above, but in accordance with the
Lie superalgebra approach of Gaiotto and Witten
\cite{Gaiotto:2008sd} we do not find such solutions.

Having satisfied the quadratic constraint by employing the above
building blocks, we now discuss the solution of the linear
constraint for the different cases with decreasing number of
supersymmetries separately. From the structure of globally
supersymmetric theories (in contrast to supergravity), it is clear
that theories with ${\cal N}$ supercharges can be seen as particular
examples of theories with lower ${\cal N}$. For this reason we will
not repeat the higher ${\cal N}$ examples when discussing the lower
${\cal N}$ theories.

\bigskip

\noindent $\bullet\ $\underline{${\cal N} \  =\  8$ superconformal
gaugings:} \\[.1cm] \noindent \ In this case the embedding tensor contains
only one irreducible component under ${\rm SO}(N)$, that is the
4-index anti-symmetric tensor $\Theta_{ab,cd} = \Theta_{[ab,cd]}$.
Therefore, one cannot use the Kronecker delta $\delta_{ab}$ within
$\Theta$.

One possibility is to make use of the special operator given in
\eqref{specialN=4} and write
\begin{equation} \Theta_{ab,cd} = g\,\epsilon_{abcd}\,,
\label{N=8et}
\end{equation} for arbitrary coupling constant $g$.
This restricts to $N=4$ and ${\rm SO}(4)$ gauging.

Another possibility is to consider a symplectic gauging and to
construct an invariant embedding tensor of the form $\Theta_{ab,cd}
\sim \Omega_{[ab}\Omega_{cd]}$. However, according to eq.~\eqref{po}
this is not an ${\rm Sp}(N)$ projection operator. Therefore, the
quadratic constraint will not be satisfied and one cannot consider
this possibility. We conclude that for ${\cal N}=8$ one can only
gauge $SO(4)$ or multiple copies thereof\footnote{Note that the gauging of a $G_2\subset {\rm SO}(7)$ subgroup in the ${\cal N}=8$ case is not possible because, although $C_{abcd}$ is a totally anti-symmetric invariant tensor of $G_2$, and thereby satisfying the linear constraint \eqref{linear8}, it does not projects onto $G_2$. On the other hand, the combination $\delta_{a[c}\delta_{d]b} + \frac14\,C_{abcd}$ does project onto $G_2$ but it does {\sl not} satisfy the ${\cal N}=8$ linear constraint \cite{Bergshoeff:2008cz}.}.

\bigskip

 \noindent $\bullet\ $\underline{${\cal N}  \ =\ 6$ superconformal gaugings:} \\[.1cm] \noindent \ In this case  we are dealing with an
embedding tensor $\Theta_a{}^b{}_{,\, c}{}^d$ that satisfies the
linear constraint \eqref{linearc}. Since the embedding tensor has
both upper and lower indices  we can use the invariant Kronecker
delta $\delta_a{}^b$ to build expressions for $\Theta$. This does
not restrict to particular values of $N$. That is the basic reason
why for ${\cal N}=6$ one can obtain gaugings for arbitrary $N$
\cite{Bergshoeff:2008ix}.

The easiest way to find a solution that satisfies the linear
constraint is to take
\begin{eqnarray}
  \Theta_a{}^b{}_{,c}{}^d &=&
  g\,\delta_{[a}{}^{[d}\,\delta_{c]}{}^{b]}\ =\ {g\over 2}\,
  \big(\delta_c{}^b\,\delta_a{}^d - {1\over
  N}\,\delta_a{}^b\,\delta_c{}^d\big)\ -\ {(N-1)\over N}{g \over
  2}\,\delta_a{}^b\,\delta_c{}^d \,,
  \label{U-singlet}
\end{eqnarray}
for arbitrary coupling constant $g$. The singlet operator
becomes a ${\rm U}(1)$ projection operator. For $N>1$ this picks out
all generators of ${\rm U}(N)$ and leads to a gauging of the full
${\rm U}(N)$ group. Note that, in order to satisfy the linear
constraint \eqref{linearc}, we must take a specific combination of
the ${\rm SU}(N)$ and ${\rm U}(1)$ operators. By taking multiple
copies thereof one obtains vector models with ${\rm U}(m_1)\times
{\rm U}(m_2) \times \cdots$ gauging, where $m_1 + m_2 + \ldots = N$.

We next consider a matrix model describing the embedding
 ${\rm U}(m)\times {\rm U}(n)\subset {\rm U}(N=mn)$ such that
the scalars transform in the bi-fundamental representation $(m,n)$.
We first try an embedding tensor that contains products of adjoints
with singlets. However, one finds that one can not satisfy the
linear constraint \eqref{linearc} with this Ansatz. For this we need
to add a common ${\rm U}(1)$ factor that acts on both factors. We
thus obtain
\begin{eqnarray}
{\Theta}_{(i,\bar i)}{}^{(k,\bar k)}{}_{,\,(j,\bar j)}{}^{(l, \bar
l)}  &=& g\, \delta_{\bar i}{}^{\bar k}\delta_{\bar j}{}^{\bar
l}\big(\delta_i{}^l \delta_j{}^k -{1\over
m}\delta_i{}^k\delta_j{}^l\big) - g \delta_j{}^l \delta_i{}^k \big(
\delta_{\bar j}{}^{\bar k}\delta_{\bar i}{}^{\bar l} - {1\over n}
\delta_{\bar i}{}^{\bar k}  \delta_{\bar j}{}^{\bar
l}\big)\nonumber\\
&&\ - \ {(m-n)\over mn}g\,\delta_i{}^k\delta_j{}^l \delta_{\bar
i}{}^{\bar k}\delta_{\bar j}{}^{\bar l}\,, \label{U-matrix}
\end{eqnarray}
for arbitrary coupling constant $g$. We deduce that the unitary
matrix model describes an ${\rm SU}(m)\times {\rm SU}(n)\times {\rm
U}(1)$ gauging, corresponding to the ${\rm U}(m|n)$ model of
\cite{Hosomichi:2008jb}. For $m=n$, in which case the ${\rm U}(1)$
factor vanishes \cite{Schnabl:2008wj}, this is the ABJM model of
\cite{Aharony:2008ug} .

Finally, we consider symplectic gaugings. Note that we can now raise
and lower indices using the symplectic tensor. We first try an
embedding tensor that only contains the adjoint of ${\rm Sp}(n)$.
However, this does not satisfy the linear constraint \eqref{linearc}
and we must add an additional ${\rm U}(1)$ factor:
\begin{equation}
\Theta_{ab,cd}\ =\ g\,\Omega_{ab}\Omega_{cd} \ -\ g\,\big(
\Omega_{ca}\Omega_{bd}\ + \ \Omega_{cb}\Omega_{ad}\big)\,,
\label{N=6et}
\end{equation}
where the first term on the right-hand-side corresponds to the ${\rm
U}(1)$ gauging and where the term between round brackets corresponds
to the ${\rm Sp}(n)$ gauging. This is precisely the so-called ${\rm
OSp}(2|n)$ model of \cite{Hosomichi:2008jb}.

\bigskip

\noindent $\bullet\ $\underline{${\cal N} \ =\ 5$ superconformal gaugings:} \\[.1cm] \noindent The global symmetry
group for ${\cal N}=5$ is ${\rm Sp}(N)$. We first try to gauge the
full ${\rm Sp}(N)$ using the Ansatz
 \bea
  \Theta_{ab,cd} \ = \ g \; \Omega_{c(a}\,\Omega_{b)d} \,,
 \eea
with arbitrary coupling constant $g$. This indeed solves the linear
constraint \eqref{linear5} and leads to a vector model with ${\rm
Sp}(N)$ gauging. Similarly, one can take multiple copies thereof
with ${\rm Sp}(N_1) \times {\rm Sp}(N_2) \times \cdots $ gauging
with $N_1+N_2+\ldots =N$.

It turns out that the vector model is a special case of a matrix
model with ${\rm SO}(m) \times {\rm Sp}(n)$ gauging. The
corresponding embedding tensor solving the linear constraint is
given by \bea \label{expressionN=5}
  \Theta_{(i\bar{i})(j\bar{j}),(k\bar{k})(l\bar{l})} \ = \
  g\left(\delta_{k[i}\;\delta_{j]l}\;\Omega_{\bar{i}\bar{j}}\;\Omega_{\bar{k}\bar{l}}
  +\;\delta_{ij}\;\delta_{kl}\;\Omega_{\bar{k}(\bar{i}}\;\Omega_{\bar{j})\bar{l}}\right)\;,
 \eea
for arbitrary coupling constant $g$. This is precisely the so-called
${\rm OSp}(m|n)$ model of \cite{Hosomichi:2008jb}. Note that the
relative strength between the $\text{SO}(m)$ and $\text{Sp}(n)$
terms is fixed by the linear constraint \eqref{linear5}. Matrix
models with ${\rm SO}(m)\times {\rm SO}(n)$ or  ${\rm Sp}(m)\times
{\rm Sp}(n)$ gauging cannot be constructed simply because one cannot
embed these in the global symmetry group ${\rm Sp}(N)$.

To construct further solutions of the constraint \eqref{linear5}, we use a method that exploits a link with certain type of superalgebras, as observed in \cite{Gaiotto:2008sd}. Accordingly, we look for a superalgebra with (anti-)commutation rules of the form
\begin{equation} \{Q_a,Q_b\} = (t^\alpha)_{ab} T_\alpha\ ,\quad\quad
[T_\alpha,Q_a]=\eta_{\alpha \beta} (t^\beta)_{ab} Q^b\ ,
\label{Jacobi}
\end{equation}
where $\eta_{\alpha\beta}$ is the Cartan-Killing metric. Then, an embedding tensor defined as
\begin{equation}
\Theta_{ab,cd} = (t^\alpha)_{ab}
(t^\beta)_{cd}\,\eta_{\alpha\beta}\ ,
\end{equation}
is guaranteed to satisfy the linear constraint \eqref{linear5} as a consequence of the Jacobi identity $[\{Q_a,Q_b\},Q_c] + {\rm perms} = 0$. Applying this to the exceptional Lie superalgebras\footnote{This method applied to the Lie superalgebra ${\rm U}(m|n)$ (denoted by $spl (m,n)$ in \cite{Rittenberg}) gives the ${\cal N}=6$ and ${\cal N}=8$ solutions, and the superalgebra $OSp(m|n)$ gives the ${\cal N}=5$ solutions already discussed above. There exist two other classes of Lie superalgebras, referred to as ``strange superalgebras'' in \cite{Frappat:1996pb} and denoted by $P(n)$ and $Q(n)$. However, as the structure constants of these algebras do not fit the pattern exhibited in \eqref{Jacobi}, the associated Jacobi identities do not correspond to the constraints on the embedding tensor. As such, these algebras do not give new solutions.} as presented in \cite{Rittenberg} in a convenient notation, we find the following three additional solutions to the constraints of the embedding tensor for ${\cal N}=5$.

In the case of the Lie superalgebra $F(4)$, the embedding tensor
reads (where $i,j,..$ refer to the spinor representation $\bf 8$ of
${\rm SO}(7)$ and $\alpha, \beta, ..$ denote an ${\rm SU}(2)$
doublet)
\begin{align} \label{excF}
 \Theta_{i\alpha\, j\beta , k\gamma\,l\delta} = \tfrac{1}{12}
\Gamma^{mn}_{ij}\Gamma^{mn}_{kl}\epsilon_{\alpha\beta}\epsilon_{\gamma\delta}
+
\delta_{ij}\delta_{kl}\epsilon_{\gamma(\alpha}\epsilon_{\beta)\delta}\,,
\end{align}
with $SO(7)$ Gamma-matrices $\Gamma^m$. This provides a solution to
the linear constraint \eqref{linear5} and gives rise to a gauging of
${\rm SO}(7) \times {\rm SU}(2)$.

The second possibility corresponds to the Lie superalgebra $G(3)$.
The embedding tensor is given by (where $i,j,..$ refer to the
fundamental representation $\bf 7$ of $G_2$ and $\alpha, \beta, ..$
denote an ${\rm SU}(2)$ doublet)
\begin{align} \label{excG}
 \Theta_{i\alpha\, j\beta , k\gamma\,l\delta} =
 (\delta_{i[k}\delta_{l]j} + \tfrac14\,C_{ijkl})\,\epsilon_{\alpha\beta}\epsilon_{\gamma\delta}
 +\delta_{ij}\delta_{kl} \epsilon_{\gamma(\alpha}\epsilon_{\beta)\delta} \,,
\end{align}
where $C_{ijkl}$ is the invariant tensor of\footnote{In showing that
the structure constants of $G(3)$, which can be found in
\cite{Rittenberg}, have the required form \eqref{Jacobi}, it is
important to note that the Cartan-Killing form of $G_2$ involves the
invariant tensor $C_{ijkl}$.} $G_2$. This leads to a $G_2 \times
{\rm SU}(2)$ gauge group.

Finally, the Lie superalgebra $D(2|1;\alpha)$ (referred to as ${\rm
OSp}(4|2;\alpha)$ in \cite{Rittenberg}) gives a deformation of the
${\rm SO}(4) \times {\rm Sp}(1)$ gauging with embedding tensor
 \begin{align} \label{excO}
 \Theta_{i\alpha\, j\beta , k\gamma\,l\delta} =
(\delta_{i[k}\delta_{l]j} +
\gamma/2\,\epsilon_{ijkl})\,\epsilon_{\alpha\beta}\epsilon_{\gamma\delta}
+\delta_{ij}\delta_{kl}
\epsilon_{\gamma(\alpha}\epsilon_{\beta)\delta}\,,
 \end{align}
with $i,j,=1, .., 4$ of $SO(4)$ and $\alpha, \beta=1,2$ of $SU(2)$.
This example corresponds to a deformation of the gauging of ${\rm
SO}(4) \times {\rm Sp}(1)$ in the ${\rm OSp}(4|1)$ model. In
standard notation $D(2|1,\alpha)$, $\alpha$ corresponds to the ratio
$(1+\gamma)/(1-\gamma)$ of the two coupling constants of $SO(4)$.

\bigskip

\noindent $\bullet\ $\underline{${\cal N} \ =\ 4$\ superconformal gaugings:}  \\[.1cm] \noindent \ A noteworthy feature
of the case of four supersymmetries is that there is a direct
product structure. Each factor has an R-symmetry ${\rm Sp}(1)$ and a
global symmetry group ${\rm Sp}(N)$. To distinguish the first sector
from the second, so-called ``twisted'' sector, we use $a$ indices
for ${\rm Sp}(N)$ and $a'$ for the twisted ${\rm Sp}(N')$. We
already mentioned  that there are three kind of embedding tensors:
those with only $a$-indices, those with only twisted $a'$-indices
and mixed embedding tensors with $a$ and $a'$ indices.

Restricting first to the untwisted sector, the set of possible
models coincides with those described above for $\cN \geq 5$. The
reason is that the $\cN = 4$ linear constraint coincides with that
of $\cN = 5$. Hence for every solution with $\cN \geq 5$ there is a
corresponding solution with $\cN = 4$. The two classes with  ${\rm
SO}(m)\times {\rm Sp}(n)$ and ${\rm SU}(m) \times {\rm SU}(n) \times
{\rm U}(1)$ were first described by \cite{Gaiotto:2008sd}. In
addition to these two regular classes, the three exceptional cases
that occurred in $\cN = 5$ also make their appearance in $\cN = 4$.
The expressions for the embedding tensor are identical to their $\cN
= 5$ counterparts.

The situation changes if we also include the twisted sector
\cite{Hosomichi:2008jd}. First of all, a relatively trivial
possibility is to include this without coupling to the untwisted
sector. This allows for additional gaugings parametrised by
$\Theta_{a'b',c'd'}$, which also has to be of one of the above
forms. More interesting is the possibility to couple to two sectors,
using the off-diagonal components $\Theta_{ab,c'd'} =
\Theta_{c'd',ab}$. It is impossible to excite this component for
generic gaugings in the untwisted and twisted sector; it can easily
be seen that there are no possible terms with the correct symmetry
properties. Indeed, an identification has to be made between the
gaugings in the two sectors, as we will now illustrate.

\renewcommand{\arraystretch}{1.4}
\begin{table}[ht]
\begin{center}
$
\begin{array}{||c||c|c||c||}
\hline \mN &  \Theta & {\rm gauge~group} & {\rm Lie~superalgebra} \\
\hline \hline
4,8 & \eqref{N=8et} & {\rm SU}(2) \times {\rm SU}(2)  & {\rm U}(2|2) \\[1mm]
\hline \hline 4,6 & \eqref{U-matrix} & {\rm SU}(m) \times {\rm
SU}(n) \times {\rm U}(1) &
{\rm U}(m|n) \\[1mm] \hline
4,6  & \eqref{N=6et} & {\rm SO}(2) \times {\rm Sp}(n) & {\rm
OSp}(2|n)
\\[1mm] \hline \hline
4,5 & \eqref{expressionN=5} & {\rm SO}(m) \times {\rm Sp}(n) & {\rm OSp}(m|n) \\[1mm] \hline
4,5  & \eqref{excF} & {\rm SO}(7) \times {\rm SU}(2) & F(4) \\[1mm]\hline
4,5  & \eqref{excG} & G_2 \times {\rm SU}(2) & G(3) \\[1mm]\hline
4,5  & \eqref{excO} & {\rm SO}(4) \times {\rm Sp}(1) & D(2|1;\alpha)
\\[1mm]\hline
\end{array}$
\caption{The equation number of the embedding tensor and gauge group
of different superconformal models for $4 \leq \cN \leq 8$ and the
associated Lie superalgebra. In the second  entry, when $m=n$ the $U(1)$
factor drops out. For $\cN = 4$ we only give the
untwisted models; when including the twisted sector non-trivial
couplings such as \eqref{offdiag1} or \eqref{offdiag2} can also be
introduced. For $\cN \leq 3$, there are no restrictions on the gauge
group. All these models also have the superconformal symmetry based
on the Lie superalgebra ${\rm OSp}({\cal N}|4)$.}
\label{tab:gaugings}
\end{center}
\end{table}
\renewcommand{\arraystretch}{1}

For concreteness we will specify to an ${\rm SO}(m)\times {\rm
Sp}(n)$ and ${\rm SO}(m')\times {\rm Sp}(n')$ gauging in both
sectors, respectively. Upon identification of the two orthogonal
sectors, i.e.~${\rm SO}(m) \simeq {\rm SO}(m')$, one needs to
include an off-diagonal term (where $a=\{ i, \bar i\}$ and $a' =\{
i', \bar i' \}$ and $i \simeq i'$)
 \begin{align}
  \Theta_{ab,c'd'} =  g \,\delta_{k'[i}\;\delta_{j]l'}\;\Omega_{\bar{i}\bar{j}}\;\Omega_{\bar{k}\bar{l}} \,.
 \label{offdiag1}
 \end{align}
This corresponds to a gauge group ${\rm Sp}(n) \times {\rm SO}(m)
\times {\rm Sp}(n')$, where the (twisted) hypermultiplets are in the
bifundamental of the first (last) two factors. Similarly, upon
identification of the two symplectic sectors, i.e.~${\rm Sp}(n)
\simeq {\rm Sp}(n')$, the following off-diagonal term has to be
included (where $a=\{ i, \bar i\}$ and $a' =\{ i', \bar i' \}$ and
$\bar i \simeq \bar i'$)
 \begin{align}
   \Theta_{ab,c'd'} =  g \;\delta_{ij}\;\delta_{k'l'}\;\Omega_{\bar{k}'(\bar{i}}\;\Omega_{\bar{j})\bar{l}'} \;.
 \label{offdiag2}
 \end{align}
In this case the gauge group is ${\rm SO}(m) \times {\rm Sp}(n)
\times {\rm SO}(m')$ where again  the (twisted) hypermultiplets are
in the bifundamental of the first (last) two factors. By subsequent
applications of this construction one can obtain an (in)finite ${\rm
SO}(m_1) \times {\rm Sp}(n_1) \times {\rm SO}(m_2) \times {\rm
Sp}(n_2) \times \cdots$ gauge group \cite{Hosomichi:2008jd}.

A distinct possibility is to identify {\it both} gauge groups,
i.e.~${\rm SO}(m) \simeq {\rm SO}(m')$ and ${\rm Sp}(n) \simeq {\rm
Sp}(n')$. In this case one needs the off-diagonal component to
consist of both terms discussed above:
 \begin{align}
  \Theta_{ab,c'd'} =  g \,\delta_{k'[i}\;\delta_{j]l'}\;\Omega_{\bar{i}\bar{j}}\;\Omega_{\bar{k}\bar{l}} + g \;\delta_{ij}\;\delta_{k'l'}\;\Omega_{\bar{k}'(\bar{i}}\;\Omega_{\bar{j})\bar{l}'} \;.
 \end{align}
Note that this leads to $\Theta_{ab,cd} = \Theta_{ab,c'd'} =
\Theta_{a'b',c'd'}$. In this case the untwisted and twisted
hypermultiplets naturally combine into $\cN = 5$ multiplets and one
finds supersymmetry enhancement \cite{Hosomichi:2008jb}, in this
case to the solution \eqref{expressionN=5}.

A similar story holds for the ${\rm SU}(m) \times {\rm SU}(n) \times
{\rm U}(1)$ gauging, where one can also employ the twisted sector to
obtain a chain of unitary gauge groups \cite{Hosomichi:2008jd}. In
addition, the identification of both unitary groups in the untwisted
and twisted sector leads to  $\cN = 6$ supersymmetry
\cite{Hosomichi:2008jb}, as the embedding tensor automatically
satisfies the corresponding linear constraint \eqref{linearc}.
Finally, one can construct couplings between the untwisted and
twisted sector in the case that these are given by one of the three
exceptional cases.

\bigskip

\noindent $\bullet\ $\underline{${\cal N} \leq 3$ superconformal gaugings:}  \\[.1cm] \noindent \ From the gauged
supergravity models with ${\cal N}=1,2$ or $3$ supersymmetries it
follows that there is {\sl no} linear constraint. Therefore, there
is no restriction on the gauge group and matter content.
Furthermore, superconformal field theories with polynomial
interactions not based on a gauging, and hence without Chern-Simons
terms, are known to exist for ${\cal N}=1, 2$ \cite{Nicolai:1988ek,
Fronsdal:1981gq, Sokatchev}. We expect them to arise in the global
limit of ${\cal N}=1,2$ supergravities with interactions that do not
follow from gauging \cite{deWit:2003ja}. In fact, both type of
interactions, namely those which are tied to gauging and the others
which are independent of gauging, can simultaneously arise in the
${\cal N}=1,2$ superconformal field theories.

\noindent This finishes our discussion of how the constraints are
solved for different values of ${\cal N}$. For the convenience of
the reader the different possibilities are summarised in table
\ref{tab:gaugings}.

\section{An example: the ${\cal N}=6$ superconformal theory} \setcounter{equation}{0}

In this section we will present more details on our construction for
the specific case of $\cN = 6$. The gauged supergravity theories in
three dimensions take the general form \cite{deWit:2003ja}
 \begin{eqnarray}\label{gauged-L}
 \cL &=& \cL_0 - e V  \\ \nonumber
&&+ e \: \Big\{
\ft12 A_1^{IJ,KL}\,\bar \Psi_{\mu\,IJ}
\Gamma^{\mu\nu}\,\Psi_{\nu\,KL} +
 A_{2}{}^{a}{}_I{}^{JK}\,\bar \Psi_{\mu\,JK} \Gamma^\mu \psi_a^I +
\ft12  A_{3\hspace{0.2em}aJ}^{\hspace{0.6em}bI}\, \bar \psi^a_I
\psi_b^J + {\rm h.c.} \Big\} \;,
 \end{eqnarray}
where ${\cal L}_0$ is the part consisting of the kinetic terms, the
Chern-Simons term and quartic fermion interactions. The fermions are
the gravitini $\Psi_{\mu\;IJ}$ and the spinors $\psi_a^I$, and
$A_1,A_2,A_3$ are complicated functions of the scalar fields
belonging to the cosets \eqref{coset} summarised in table 1, and the
scalar potential $V$ consists of a sum of squares of $A_1$ and
$A_2$. Performing the conformal limit described in Section 2.2
produces the superconformal field theories in which only gauge
invariant kinetic terms for a suitable number of scalars and
spinors, along with a potential for the scalars, Yukawa terms and a
Chern-Simons term survive. In particular, the conformal truncations
of the functions $A_2$ and $A_3$ survive the limit, while $A_1$
disappears (although it does play an important role in determining
the possible massive deformations of the conformal field theory
\cite{Bergshoeff:2008ix}).

Instead of performing the steps outlined above, knowing the
structure of the terms that survive the conformal limit, we have
directly constructed the superconformal Lagrangian by the Noether
procedure. We have done so independently of the results that
appeared in the course of our work. Indeed, the ${\cal N}=6$
superconformal gauge theory was first discussed in
\cite{Aharony:2008ug}. Its ${\cal N}=2$ superspace formulation was
presented in \cite{Benna:2008zy}. The explicit form of the
supersymmetry rules have first been given in \cite{Gaiotto:2008cg,
Hosomichi:2008jb}.

The theory contains $4N$ complex scalars $X_{aI}$, where $I,J,\ldots
=(1,\ldots,4)$ and $a,b,\ldots =(1,\ldots,N)$, together with their
complex conjugates $(X_{aI})^* = X^{aI}$. They transform in the
$(4,N)$ fundamental representation of ${\rm SU}(4)\times {\rm
U}(N)$. The fermions are given by two-component Dirac spinors
$\psi_I^a$ describing $4N$ complex fermionic degrees of
freedom.\footnote{Our representation assignment for the scalars and
spinors differ from those that have appeared in the literature where
$(4,\bar N)$ and $(4,N)$, respectively, are used.}

The corresponding action can be obtained by taking the conformal
limit of ${\cal N}=6$ supergravity. In this limit the scalar
manifold reduces to a flat space, and thus in the ungauged case this
results in the free Lagrangian \cite{Bergshoeff:2008ix}
  \bea\label{free}
   {\cal L} \ = \ -\ft12\partial^{\mu}X^{aI}\partial_{\mu}X_{aI}
   +\ft12\bar{\psi}_{a}^{I}\gamma^{\mu}\partial_{\mu}\psi_{I}^{a}\;,
  \eea
which exhibits ${\cal N}=6$ superconformal symmetry.

The conformal limit in the gauged case then corresponds to a gauging
of (\ref{free}). Its action takes the general form
\begin{eqnarray}\label{lagrangianf}
   {\cal L} &=&
   -\ft12D^{\mu}X^{aI}D_{\mu}X_{aI}+\ft12\bar{\psi}_{a}^{I}\gamma^{\mu}D_{\mu}\psi_{I}^{a}\nonumber
   + {\cal L}_{Y} \\
   &&-\ft12\varepsilon^{\mu\nu\rho}A_{\mu}{}^{\alpha}\Theta_{\alpha\beta}\left(\partial_{\nu}A_{\rho}{}^{\beta}
   -\ft13\Theta_{\gamma\delta}f^{\beta\delta}{}_{\epsilon}A_{\nu}{}^{\gamma}A_{\rho}{}^{\epsilon}\right)
   -\ft{1}{3}
   A_{2}{}^{a}{}_{I}{}^{JK} {A}_{2\;a}{}^{I}{}_{JK} \;,
\end{eqnarray}
where the Yukawa couplings ${\cal L}_{Y}$ are determined by the
global limit of $A_3$. These couplings and the function $A_2$ that
determines the potential will be specified below. Note that gauge
vectors are introduced via covariant derivatives and a Chern-Simons
term. The covariant derivatives of the scalars are defined as
follows
  \bea
  \begin{split}
   D_{\mu}X_{aI} \ &= \
   \partial_{\mu}X_{aI}-\Theta_{a}{}^{b}{}_{,\,c}{}^{d}A_{\mu
   d}{}^{c}X_{bI}\;,
   \\
   D_{\mu}X^{aI} \ &= \
   \partial_{\mu}X^{aI}+\Theta_{b}{}^{a}{}_{,\,d}{} ^{c}A_{\mu
   c}{}^{d}X^{bI}\;,
  \end{split}
  \eea
and a similar definition applies to the covariant derivatives of the
spinors. Here, we have used that the ${\rm U}(N)$ gauge fields are
anti-Hermitian, $(A_{\mu\;a}{}^{b})^{*}=-A_{\mu\;b}{}^{a}$ and the
embedding tensor Hermitian, $(\Theta_{a}{}^{b}{}_{,c}{}^{d})^* =
\Theta_{b}{}^{a}{}_{,d}{}^{c}$. The Yukawa couplings read explicitly
  \bea
  \begin{split}
   {\cal L}_{Y}  \ = \  & -\ft12\Theta_{a}{}^{b}{}_{,\,c}{}^{d}
   X_{dI}X^{cJ}{\bar\psi}^I_b\psi^a_J + \ft{1}{4}
\Theta_{a}{}^{b}{}_{,\,c}{}^{d} X_d{}^c
   {\bar\psi}^I_b\psi^a_I  \\
   &+\ft{1}{8}\left(\epsilon_{IJKL}\Theta_{c}{}^{a}{}_{,\,d}{}^{b}X^{Ic}X^{Jd}
   {\tilde\psi}^K_a\psi^L_b\ + \ \text{h.c.}\right)\;,
  \end{split}
  \eea
where $\psi_{b}^{L}=(\psi_{L}^{b})^{\star}$ and $\tilde \psi^K_a$ is
defined in the appendix. Note that conformal invariance forbids the
occurrence of quartic fermion terms. Finally, the tensor
$A_{2}{}^{a}{}_{I}{}^{JK}$ defining the scalar potential is given by
  \begin{eqnarray}\label{A2}
   A_{2}{}^{a}{}_{I}{}^{JK}(X) &=&
   -\ft12\Theta_{b}{}^{a}{}_{,\,c}{}^{d}\left(X^{bJ}X^{cK}X_{dI}
   +\delta_{I}{}^{[J}X^{K]b}X_d{}^c\right)
   \;,
  \end{eqnarray}
where we have used the abbreviation $X_a{}^b \equiv X_{aI}X^{bI}$,
and $A_{2a}{}^I{}_{JK} \equiv (A_2{}^a{}_I{}^{JK})^*$.

The supersymmetry transformations leaving invariant
\eqref{lagrangianf} are given by
\begin{eqnarray}
  \delta X^{aI} \ &=& \ \bar{\epsilon}^{IJ}\psi_{J}^{a}\,,\nonumber\\
  \delta \psi_{I}^{a} \ &=& \
    \gamma^{\mu}D_{\mu}X^{a
    J}\epsilon_{IJ}
    +A_{2}{}^{a}{}_{I}{}^{JK}(X)\epsilon_{JK}\,,\\
   \nonumber \delta A_{\mu\hspace{0.1em}a}{}^{b} \ &=& \
   \ft12\bar{\psi}_{a}^{I}\gamma_{\mu}X^{bJ}\epsilon_{IJ}-{\rm
   h.c.}\;,
  \end{eqnarray}
where the supersymmetry parameter $\epsilon_{IJ}$ is in the real
antisymmetric representation of ${\rm SU}(4)$ and satisfies a
reality condition, see \eqref{reality}. They leave
\eqref{lagrangianf} invariant, provided the embedding tensor
satisfies the linear constraint \eqref{linearc} and the quadratic
constraints \eqref{Q6}. Note that the action takes a `universal'
form in terms of the embedding tensor in that any particular gauging
corresponds to a specific choice of $\Theta$ in \eqref{lagrangianf},
subject to the linear and quadratic constraints \eqref{linearc} and
\eqref{Q6}, respectively.

It is instructive to verify the supersymmetry of the action
corresponding to \eqref{lagrangianf}. The lowest-order supersymmetry
variation of the kinetic terms no longer vanishes due to the
non-commutativity of the covariant derivatives. Up to a total
derivative we obtain
  \bea
   \delta {\cal L}_{\rm kin} \ = \
   \ft14\Theta_{b}{}^{a}{}_{,d}{}^{c}\bar{\psi}_{a}^{I}\gamma^{\mu\nu}
   F_{\mu\nu\hspace{0.1em}c}{}^{d}X^{bJ}\epsilon_{IJ}\;.
  \eea
These variations are canceled by the  supersymmetry variation of the
gauge vectors in the Chern-Simons term. The variation of the gauge
vectors inside the covariant derivatives  gives rise to additional
contributions linear in $\Theta$. These are canceled by taking the
$\Theta$-dependent terms (parameterized by $A_2$) of
$\delta\psi^{a}_{I}$ in the variation of the fermion kinetic term
and by taking the $\Theta$-independent term in the variation of the
fermions in the Yukawa terms. This cancelation takes place provided
the linear constraint \eqref{linearc} on $\Theta$ holds.

We next consider the variations quadratic in $\Theta$.  The
variation of the Yukawa couplings leads to two types of terms,
$\bar{\psi}_{a}^{I}\epsilon_{JK}$ and
$\bar{\psi}_{a}^{I}\epsilon_{IJ}$, i.e.~with uncontracted or
contracted ${\rm SU}(4)$ indices. The former terms cancel among
themselves, which can be proven upon using linear combinations of
the quadratic constraints \eqref{Q6} with different index
permutations. Similarly, the latter terms cancel against the
variations of the scalar potential.

Let us finally note that the supersymmetric action for all ${\cal
N}\leq 5$ take the same universal form as \eqref{lagrangianf}, in
which the Yukawa couplings and scalar potential are parameterized by
the tensors $A_2$ and $A_3$ that can be obtained from the
supergravity models of \cite{deWit:2003ja} as in
\cite{Bergshoeff:2008ix}. As mentioned earlier, in the case of
${\cal N}=1,2$  special features arise due to the fact that the
supergravity theories admit interactions that are independent
of gauging as well \cite{deWit:2003ja}.

We also note that all of the actions described in this paper, for
which ${\cal N}$ ordinary supersymmetries are explicitly given, are
also invariant under special supersymmetries, dilatation and
conformal boosts, in addition to the manifest $R$-symmetry groups
$H_R$ (see table 2). Altogether these form the superconformal group
${\rm OSp}({\cal N}|4)$. The symmetry transformations other than the
ordinary supersymmetries are not repeated here (see, for example,
\cite{Bandres:2008vf}).

\section{Discussion} \setcounter{equation}{0}

In this work we used the three-dimensional gauged supergravity
models of  \cite{Nicolai:2001ac, deWit:2003ja} to obtain information
about superconformal gauge theories in three dimensions for an
arbitrary number ${\cal N}\le 8$  of supersymmetries. The embedding
tensor characterizing the superconformal theory satisfies a linear
and a quadratic constraint. For each solution of these constraints
one obtains a consistent gauging.  We solved the constraints using a
simple tensor analysis and presented the gauge groups and matter
content of the different superconformal theories. We find all the
superconformal theories that occur in the recent classification of
\cite{Gaiotto:2008sd,Hosomichi:2008jd,Hosomichi:2008jb,Schnabl:2008wj}.
On top of that we find three new superconformal theories with ${\cal
N}=4,5$ supersymmetry. These latter cases are suggested by the
relation with the Lie superalgebras \cite{Gaiotto:2008sd}.

 The supergravity approach allows to construct  non-conformal gaugings
and deformations as well  \cite{Bergshoeff:2008ix}. These include
 (1) massive deformations of the superconformal theories and (2)
standard Yang-Mills
 gauge theories. The massive deformations occur in two types: (1a)
 scalar massive deformations and (1b) vector massive deformation.
 In the former case one introduces mass parameters for a number of
scalar fields.
 In the latter case one gauges translations corresponding to a number
of scalar fields. This requires the introduction of new gauge vector
fields, with a corresponding Chern-Simons term. In the gauge where
the scalars are vanishing, the vector fields obtain a mass term in
addition to their Chern-Simons term.
 By taking a non-conformal limit of gauged
 supergravity it can be shown that the scalar and vector mass
parameters of the ${\cal N}=8$
 superconformal theory occur in the following representations of the
R-symmetry group ${\rm SO}(8)$
 \cite{Bergshoeff:2008ix}\,:
 \begin{equation}
 \text{scalar masses}\,: \ \ \ {\bf 35}_s\,,\hskip 2truecm
 \text{vector masses}\,: \ \ \ {\bf 35}_v\,.
 \end{equation}
Decomposing into ${\rm SU}(4)\times {\rm U}(1)$ and projecting onto
${\rm U}(1)$ singlets suggests that the ${\cal N}=6$ superconformal
theory can be deformed by the following representations of the ${\rm
SU}(4)$ R-symmetry group\,:
\begin{equation}
 \text{scalar masses}\,: \ \ \ {\bf 15} \,,\hskip 2truecm
 \text{vector masses}\,: \ \ \ {\bf 15}\,.
 \end{equation}
Similarly, this leads one to expect the following representations of
the $Sp(2)$ R-symmetry group for $\cN = 5$:
 \begin{equation}
 \text{scalar masses}\,: \ \ \ {\bf 5} \,,\hskip 2truecm
 \text{vector masses}\,: \ \ \ {\bf 5}\,.
 \end{equation}
The presence of a scalar mass term breaks the R-symmetry group to
$SO(4) \times SO(\cN - 4)$ \cite{Gomis:2008cv,Hosomichi:2008qk,
Hosomichi:2008jb,Gomis:2008vc}. Continuing to lower ${\cal N}$ one
could in this way classify massive deformations of all
superconformal theories. It would be interesting to construct these
deformations by taking the non-conformal limit of the gauged
supergravity models of
 \cite{deWit:2003ja} and study their interplay with conformal and
non-conformal gaugings.

\subsection*{Acknowledgments}

We acknowledge discussions with Joaquim Gomis, Neil Lambert, Jaemo
Park and Paul Townsend. E.A.B.~thanks the physics department of
Barcelona University for their kind support and hospitality.
D.R.~wishes to thank the Ecole Normale Sup\'erieure de Lyon for its
hospitality. E.S.~is grateful to University of Groningen, Ecole
Normale Sup\'erieure de Lyon and SISSA for hospitality during the
course of this work. This work was partially supported by the
European Commission FP6 program MRTN-CT-2004-005104EU and by the
INTAS Project 1000008-7928. In addition, the work of D.R.~has been
supported by MCYT FPA 2004-04582-C02-01 and CIRIT GC 2005SGR-00564.
The work of H.S. was supported in part by the Agence Nationale de la
Recherche (ANR). The research of E.S. is supported in part by NSF
grant PHY-0555575.

\begin{appendix}
\renewcommand{\theequation}{\Alph{section}.\arabic{equation}}

\section{Notations and Conventions} \setcounter{equation}{0}

This appendix contains information about the notation and conventions pertaining to the $\cN=6$ theory discussed in section 4.

We choose the space-time metric to be $\eta={\rm diag}(-++)$. The
gamma matrices satisfy the Clifford algebra
$\{\gamma^{\mu},\gamma^{\nu}\} =2\eta^{\mu\nu}$ and obey the
identities
  \bea
   \left(\gamma^{\mu}\right)^{\dagger} \ = \
   \gamma_{0}\gamma^{\mu}\gamma_{0}\;, \qquad
   \left(\gamma^{\mu}\right)^{T} \ = \
   -C\gamma^{\mu}C^{-1}\;, \qquad
   \left(\gamma^{\mu}\right)^{*} \ = \
   B\gamma^{\mu}B^{-1}\;,
  \eea
  where $C^T=-C$ is the charge conjugation matrix and $B=-C\gamma_0$.
  Note that $C^\dagger C=1\,,C^* = -C^{-1}$ and $B^* B=1$.
In case of $U(N)$ symmetry, we use complex notation, i.e.~the spinor
fields are two-component Dirac spinors. A single Dirac spinor
describes 2 real physical degrees of freedom. We define the Dirac
conjugate as
  \bea\label{inv1}
   \bar{\psi}^I_a \ = \ \left(\psi_I^a\right)^{\dagger}i\gamma_{0}\;,
  \eea
such that $\bar{\psi}\psi$ is a (real) Lorentz scalar. For Dirac
spinors there is a second bilinear invariant, defined by
  \bea\label{inv2}
   {\tilde \psi}_I^a\psi_J^b \ = \
i\left(\psi^{T}\right)_I^aC\psi_J^b\; \qquad {\rm and}\qquad
   {\tilde \psi}^I_a \psi^J_b\ = \
   i\left(\psi^T\right)^I_a C^{-1}\psi^J_b\;,
  \eea
where $\psi_{a}^{I} \ = \ (\psi_{I}^{a})^{\star}$. In case of
Majorana spinors, satisfying $\bar{\psi}=\psi^{T}C$, the two
invariants defined in (\ref{inv1}) and (\ref{inv2}) coincide. The
supersymmetry parameter satisfies a reality condition in order to be
compatible with ${\cal N}=6$ supersymmetry,
  \bea \label{reality}
   \left(\epsilon_{IJ}\right)^{\star} \ = \ \epsilon^{IJ} \ = \ \ft12
   B\,\varepsilon^{IJKL}\epsilon_{KL}\;.
  \eea
Using this reality constraint, the supersymmetry transformation of
the complex conjugate spinor $\psi_{a}^{I}$ for ${\cal N}=6$ reads
  \bea
   \delta_{\epsilon}\psi_a^{I} \ = \ \ft12
   B\left(\varepsilon^{IJKL}\gamma^{\mu}D_{\mu}X_{aJ}
   +\varepsilon^{KLPQ}\bar{A}_{2a}{}^{I}{}_{PQ}\right)\epsilon_{KL}\;.
  \eea

\end{appendix}

\end{document}